\documentstyle[aps,multicol,epsfig]{revtex}

\newcommand{\simle}
{\raisebox{-0.75ex}[-1.5ex]{$\;\stackrel{<}{\sim}\;$}}
\newcommand{\simge}
{\raisebox{-0.75ex}[-1.5ex]{$\;\stackrel{>}{\sim}\;$}}
\def\d{{\partial}}
\def\s{{\sigma}}
\def\e{{\epsilon}}
\def\k{{ {\bf k} }}

\def\Q{{ {\bf Q} }}

\def\w{{\omega}}
\def\a{{\alpha}}

\begin{document}
%\pagestyle{plain}

%%%%%%    TEXT START    %%%%%%
\def\runtitle{
From Kondo Effect to Fermi Liquid
}
\def\runauthor
 {Hiroshi {\sc Kontani}$^1$ and Kosaku {\sc Yamada}$^2$}

\title{
From Kondo Effect to Fermi Liquid
}

\author{
Hiroshi {\sc Kontani}$^1$ and Kosaku {\sc Yamada}$^2$
}

\address{
$^1$Department of Physics, Nagoya University,
Nagoya 464-8602. \\
$^2$Department of Physics, Kyoto University, 
Kyoto 606-8502.
}

\date{\today} 

\maketitle

\begin{abstract}
The Kondo effect has been playing an important role in strongly 
correlated electon systems. The important point is that the magnetic 
impurity in metals is a typical example of the Fermi liquid. In the 
system the local spin is conserved in the ground state and continuity 
with respect to Coulomb repulsion $U$ is satisfied. This nature is 
satisfied also in the periodic systems as far as the systems remain 
as the Fermi liquid. This property of the Fermi liquid is essential 
to understand the cuprate high-$T_{\rm c}$ superconductors (HTSC). 
On the basis of the Fermi liquid theory we develop the transport 
theory such as the resistivity and the Hall coefficient in strongly 
correlated electron systems, such as
HTSC, organic metals and heavy Fermion systems.
The significant role of the vertex corrections for total charge-
and heat-currents on the transport phenomena is explained.
By taking the effect of the current vertex corrections into account,
various typical non-Fermi-liquid-like transport phenomena 
in systems with strong magnetic and/or superconducting flucutations
are explained within the Fermi liquid theory.
\end{abstract}

%\sloppy

\begin{multicols}{2}

\begin{figure}
\begin{center}
\epsfig{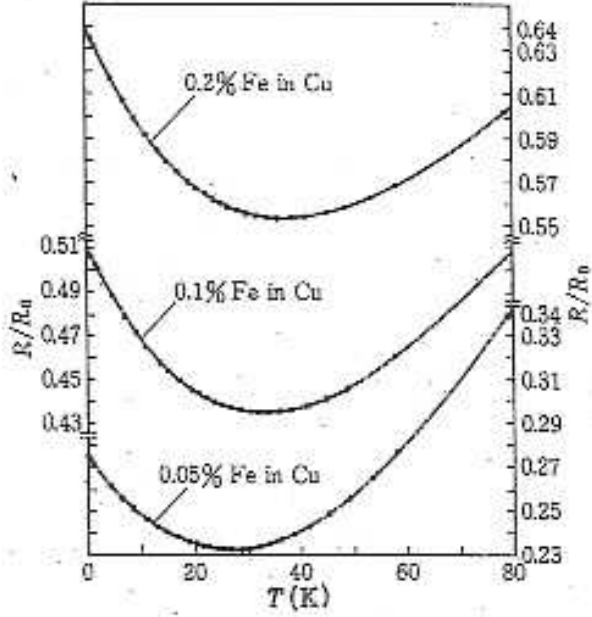}
\end{center}
\caption{
An experiment showing the resistance minimum. 
[J.P.Franck et.al.: Proc. Roy. Soc. A {\bf 263} (1961) 494.]
}
  \label{fig:K1}
\end{figure}

{\it 1.Introduction}\\
Since the discovery of Kondo Effect, the theory of electron correlation 
has been developed remarkably. In particular, the study of the strongly
 correlated electron systems such as cuprates and heavy fermions is 
actively performed. In this article we discuss the contribution of 
Kondo theory to the theory of strongly correlated electron systems. 
The key word is the singlet: The ground state of the single impurity 
in metals is the singlet where a localized spin is coupled 
antiferromagnetically with conduction electron spin. 
The resonating-valence-bond (RVB) state is a spin singlet state and 
is considered to be the basis of the cuprate high-$T_{\rm c}$ 
superconductivity (HTSC). By using Anderson's orthogonality theorem, 
we show that the Fermi liquid state is nothing but the RVB state 
in metal. As a result, we can reasonably arrive at the pairing 
theory of superconductivity.  
\\
\\
{\it 2. Anderson Hamiltonian}\\
In 1961 Anderson presented the following Hamiltonian and explained the 
appearance of a localized moment in metals by using the Hartree-Fock 
approximation. 
\cite{Anderson}\\
\begin{eqnarray}
H=\sum_{k,\sigma}\varepsilon_{k} {C_{k,\sigma}}^\dagger 
C_{k,\sigma}+\frac{1}{\sqrt{N}}\sum_{k,\sigma}(V_{kd}
C_{k\sigma}^{\dagger}d_{\sigma}+V_{dk}d_{\sigma}^{\dagger}
C_{k\sigma})\nonumber \\
+\sum_{\sigma}\varepsilon_{d}n_{d\sigma}+Un_{d\uparrow}n_{d\downarrow}.
\label{eqn:And}
\end{eqnarray}
The first term is the energy of conduction electrons and the third term 
is the d-electron part. The last term is the Coulomb repulsion between 
two d-electrons and is the only many-body interaction. 
The symmetric Anderson Hamiltonian possesses  the electron- hole 
symmetry and is easy to understand the physical meaning because of 
the suppressed charge fluctuations. By assuming the average number 
of d-electrons is unity and taking the nonmagnetic case 
$(<n_{d\sigma}>=<n_{d-\sigma}>=1/2)$ ,
\begin{eqnarray}
\varepsilon_d=-\frac{U}{2},\\
E_d=\varepsilon_d+U/2=0,
\end{eqnarray}
where $E_d$ is the averaged d-level in the nonmagnetic state. 
When one d-electron occupies the d-orbit, the energy of d-electron 
is $-U/2$ and the second d-electron occupies the level at the 
energy of $\varepsilon_d+U=U/2$. The second term of the 
Anderson Hamiltonian $(\ref{eqn:And})$ represents the mixing 
term between d- and conduction electrons. Owing to this term 
the d-electron level possesses the lifetime width $\Delta$:
\begin{equation}
\Delta=\frac{\pi}{N}\sum_k |V_{kd}|^2\delta(\varepsilon_k-E_d)
=\frac{\pi\rho|V|^2}{N}
\end{equation}
Here $\rho$ is the density of states of conduction electrons 
at the Fermi energy. Hereafter, we assume the band-width of 
conduction band is infinite and $\rho$ is constant.\\\\
{\it 2-1. Kondo effect}\\
Anderson explained the existence of magnetic moment as the 
effect of electron correlation. However, there existed a 
long standing mystery for these 30 years.
\cite{Franck}
As shown in Fig.$\ref{fig:K1}$, when we decrease the temperature, 
the residual resistivity begins to increase and the resistivity 
shows the minimum at a temperature. Kondo explained it in his 
paper published in 1964.\cite{Kondo}

He used the following s-d Hamiltonian:
\begin{eqnarray}
H=\sum_{k\sigma}\varepsilon_k C_{k\sigma}^\dagger 
C_{k\sigma}\nonumber \\
-\frac{J}{2N}\sum_{kk'\sigma\sigma'}{C_{k'\sigma'}}^\dagger 
\sigma_{\sigma'\sigma}C_{k\sigma}\cdot S \nonumber \\
=H_0+H_{sd}.
\label{eqn:sd}
\end{eqnarray}
The s-d interaction in the second term can be derived from 
Eq.$(\ref{eqn:And})$ by assuming one d-electron and $U/(\pi\Delta)$ 
is sufficiently large and by expanding with respect to mixing term. 
In the case $J$ is written as the following and possesses negative 
sign.$(J<0)$.
\begin{eqnarray}
J=-2|V|^2(\frac{-1}{\varepsilon_d}+\frac{1}{\varepsilon_d+U})
=-\frac{8|V|^2}{U}.\end{eqnarray}
In the last equation we have assumed the symmetric case, 
$\varepsilon_d=-U/2$. 
We calculate 
the T-matrix by using the s-d interaction of  Eq.$(\ref{eqn:sd})$  
as perturbation. By calculating up to the second Born term, 
we obtain the following temperature dependent term,
\begin{eqnarray}
T(\varepsilon)=-\frac{J}{2N}(S\cdot\sigma)[1+\frac{J}{2N}
\sum_{k}\frac{1-2f(\varepsilon_{k})}{\epsilon+i\eta-\varepsilon_k}]
\label{eqn:BT}
\end{eqnarray}
Here $f(\varepsilon)$ is the Fermi distribution function. 
By using the T-matrix, we obtain the electrical resistivity,
\begin{eqnarray}
\label{eqn:Kond}
R=R_B[1+\frac{2J\rho}{N}\log{\frac{k_BT}{D}}],\\
X=\frac{J\rho}{N}\log{\frac{k_BT}{D}},
\end{eqnarray}
where $R_B$ is the resistivity obtained from the first term 
of Eq.$(\ref{eqn:BT})$, that is the resistivity by the 
Born approximation. $D$ is the band-width of conduction band, 
and $k_B$ is the Boltzmann constant. 
Equation $(\ref{eqn:Kond})$ represents the logarithmic 
increase of resistivity with decreasing temperature. 
Thus, the problem of resistance minimum has been solved clearly.
%%%%%%%%%%%%%%%%%%%%%%%%%%%%%%%%%%%%%%%%%%%%%%%%%%%%%%%
%%%%%%%%%%%%%%%%%%%%%%%%%%%%%%%%%%%%%%%%%%%%%%%%%%%%%%

 However, there remains another problem, the so called Kondo 
problem; problem of the logarithmic divergence. The perturbation 
theory cannot be used when the second term of Eq.$(\ref{eqn:Kond})$ 
approaches to unity. In that case we need higher order terms. 
In $X$, the coupling constant$|\frac{J\rho}{N}|<<1$ is 
a small parameter and $X$ becomes unity when $T=T_K$:
\begin{eqnarray}
k_BT_K=D\exp[\frac{-N}{|J|\rho}]
\label{eqn:TK}
\end{eqnarray}
From this result, we have to calculate the higher order terms 
in the vicinity of $T_K$. This calculation was done by Abrikosov 
and was published in 1965.\cite{Abrikosov}. 
The most divergent terms $[J\rho/N\log(k_BT/D)]^n$ are summed 
up to the infinite order as
\begin{eqnarray}
R=R_B[1-\frac{J\rho}{N}\log(\frac{k_BT}{D})]^{-2}
\end{eqnarray}
This expression also diverges at $T=T_K$.

On the other hand, Yosida and Okiji calculated the spin 
susceptibility due to the localized spin and obtained the 
following result:\cite{YosidaOkiji}
\begin{eqnarray}
\chi_{imp}=\frac{C}{T}[1+\frac{J\rho}{N}
\frac{1}{1-\frac{J\rho}{N}\log(\frac{k_BT}{D})}],\\
C=\frac{(g\mu_B)^2S(S+1)}{3k_B},
\end{eqnarray}
where $C$ is the Curie constant and $g$ and $\mu_B$ are 
the $g$-value and Bohr magneton. This equation shows that 
the effective Curie constant becomes small with decreasing 
temperature from high temperature above $T_K$. This means 
that the localized spin becomes small toward $T_K$.

From this result Yosida proposed a singlet ground state:\cite{Yosida}
\begin{eqnarray}
\Psi_{singlet}=\frac{1}{\sqrt{2}}[\varphi_{\alpha}\chi_{\alpha}-\varphi_{\beta}\chi_{\beta}].
\label{eqn:Yosi}
\end{eqnarray}
Here $\chi_{\alpha}$ and $\chi_{\beta}$ are the wave-functions 
of localized spin and that for up and down spins, respectively. 
The wave-functions $\varphi_{\alpha}$,$\varphi_{\beta}$ are the
 wave-functions of conduction electrons associated to the 
localized spin states. $\varphi_{\alpha}$ has the down spin and 
$\varphi_{\beta}$ has up spin; they compose the spin singlet 
state with $\chi_{\alpha}$ and $\chi_{\beta}$.
%%%%%%%%%%%%%%%%%%%%%%%%%%%%%%%%%%%%%%%%%%%%%%%%%%%%%%%
\begin{figure}
\begin{center}
\epsfig{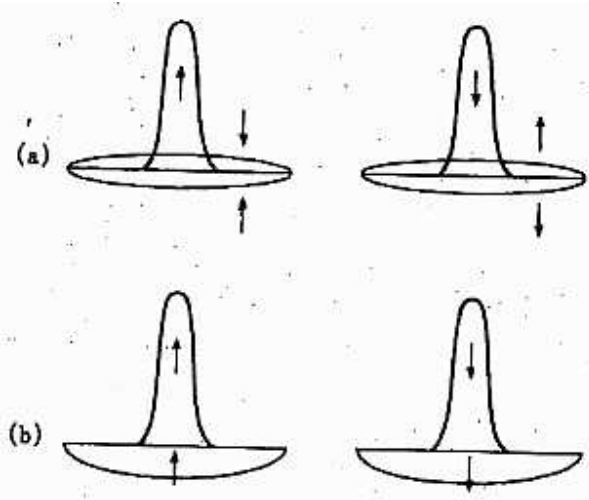}
\end{center}
\caption{
(a) conduction electron distribution around d-electron with up 
and down spins. (b)electron distribution shifted by 1/2 for 
both spins from (a). The spin is conserved if one sums up spins 
of d-electron and neighboring electrons.}
[Reference \cite{KYamada1}]
  \label{fig:S1}
\end{figure}
%%%%%%%%%%%%%%%%%%%%%%%%%%%%%%%%%%%%%%%%%%%%%%%%%%%%%%
We show the electronic states in Fig.$\ref{fig:S1}$

Substituting this wave-function into the s-d Hamiltonian 
$(\ref{eqn:sd})$ and solving the Schr$\ddot{o}$dinger equation, we obtain 
\begin{eqnarray}
H\Psi_{singlet}=E\Psi_{singlet},\\
E= E_D+\tilde{E}\\
\tilde{E}=-k_BT_K=-D\exp[-\frac{N}{\rho|J|}],
\end{eqnarray}
where $E_D$ is the normal term in the ground state energy 
which can be obtained by the perturbation expansion with respect 
to the s-d interaction, starting from the doublet state composed of 
the Fermi sphere and a localized spin. The singular term $\tilde{E}$ 
is the binding energy in the spin singlet state. This result shows 
that the system gains the energy equal to $T_K$.

The energy gain of the spin singlet state can be naturally understood 
as follows. The localized up-spin state $\chi_{\alpha}$ and the 
localized down-spin state $\chi_{\beta}$ are degenerate with each 
other. We can gain the binding energy by combining the two states 
through the transverse component of s-d exchange interaction 
$J_{\bot}$, and lifting the degeneracy. In order to realize this 
procedure, it is necessary for the following matrix elements to 
exist between the two components in the singlet ground-state 
wave-functions:
\begin{eqnarray}
<\varphi_{\alpha}\chi_{\alpha}|-\frac{J_{\bot}}{N}
\sum_{k,k'}{C_{k'\downarrow}}^{\dagger}C_{k\uparrow}S_{+}
|\varphi_{\beta}\chi_{\beta}>\neq 0 ,
\label{eqn:Jtra1}
\end{eqnarray}
\begin{eqnarray}
<\varphi_{\beta}\chi_{\beta}|-\frac{J_{\bot}}{N}\sum_{k,k'}
{C_{k'\uparrow}}^{\dagger}C_{k\downarrow}S_{-}|\varphi_{\alpha}
\chi_{\alpha}>\neq 0 .
\label{eqn:Jtra2}
\end{eqnarray}
Furthermore, for the matrix elements not to vanish, the wave-functions 
of conduction electrons should satisfy Anderson's orthogonality theorem.

The Anderson's orthogonality theorem is the following.\cite{Antyoku}.
Let $\varphi_{1}$ and $\varphi_{2}$ be two wave-functions of 
conduction electrons corresponding to two different ground states.
 The phase shifts at the Fermi energy for $\varphi_{1}$ 
and $\varphi_{2}$ are denoted as $\delta_1$ and $\delta_2$, respectively. 
Then, the overlap integral between the two ground states is given by 
\begin{eqnarray}
|<\varphi_1|\varphi_2>|\simeq N^{-(\delta_1-\delta_2)^2/(2{\pi}^2)}.
\end{eqnarray}
This is the orthogonality theorem by Anderson. $N$ is the number of 
related conduction electrons, which is a macroscopic number. 
As a result, when $\delta_1$ and $\delta_2$ do not coincide 
with each other, the two wave-functions are orthogonal and the 
overlap integral vanishes. 
%Here, $n_1=\delta_1/\pi, n_2=\delta_2/\pi$ denote the local electron numbers. 
According to the Friedel sum rule, the phase shifts $\delta_1$ and $\delta_2$
are related to the local electron numbers $n_1$ and $n_2$
as $n_1=\delta_1/\pi$ and $n_2=\delta_2/\pi$. 
Therefore, for the overlap integral to be finite, the conservation law, 
$n_1=n_2$, should be satisfied. This theorem holds also for the 
many-body matrix element. For the matrix elements, (\ref{eqn:Jtra1}) 
and (\ref{eqn:Jtra2}), not to vanish, the local electron numbers 
should satisfy the following condition. Denoting the local conduction 
electron number with $\sigma$ spin for localized up(down) spin 
as ${n_{\alpha}}^{\sigma}({n_{\beta}}^{\sigma})$, we obtain 
\begin{eqnarray}
n_{\alpha}^{\uparrow}=n_{\beta}^{\uparrow}-1 ,
\label{eqn:KonAn}\\
n_{\alpha}^{\downarrow}=n_{\beta}^{\downarrow}+1 ,\\
n_{\alpha}^{\uparrow}+n_{\alpha}^{\downarrow}=0 ,\\
n_{\beta}^{\sigma}=n_{\alpha}^{-\sigma}.
\end{eqnarray}
Here we have used that $C_{0\sigma}^{\dagger}=\frac{1}{\sqrt{N}}
\sum_kC_{k\sigma}^{\dagger}$ and $C_{0\sigma}=\frac{1}{\sqrt{N}}
\sum_kC_{k\sigma}$ creates and destroys one electron at the origin, 
respectively. From the above condition, we can determine the local 
electron numbers as\cite{YosidaYoshimori}
\begin{eqnarray}
n_{\alpha}^{\downarrow}=n_{\beta}^{\uparrow}=\frac{1}{2},  \\
 n_{\alpha}^{\uparrow}=n_{\beta}^{\downarrow}=-\frac{1}{2} .
\label{eqn:KonA}
\end{eqnarray}
This result means, as shown in Fig.$\ref{fig:S1}$, that the 
up-spin component of localized spin is combined with 
a half conduction electron 
with down-spin and a half up-spin conduction hole. 
The down-spin component of the 
localized spin is combined with 
a half up-spin conduction electron. 
Thus the singlet ground state is realized.
The local number of electrons $\pm 1/2$ corresponds to the 
phase shift of $\pm \pi/2$, and $|\sin(\delta)|=1$ gives the 
resistivity at the unitarity limit. This value corresponds to 
the nonmagnetic case of manganese.

Let us consider the case of the Anderson Hamiltonian given by 
Eq.$(\ref{eqn:And})$.\cite{KYamada1}. In this case the ground 
state wave-function is written by a linear combination of the 
following four components of d-electron state. 
%They are the states with d-electron of $0$,$\uparrow, \downarrow,2$.
\begin{eqnarray}
\Psi_g=A_0\varphi_0+A_{\uparrow}d_{\uparrow}^{\dagger}
\varphi_{\uparrow}+A_{\downarrow}d_{\downarrow}^{\dagger}
\varphi_{\downarrow}+A_2d_{\uparrow}^{\dagger}
d_{\downarrow}^{\dagger}\varphi_2 ,
\label{eqn:wavfun}
\end{eqnarray}
where $\varphi_0$, $\varphi_{\uparrow}$, $\varphi_{\downarrow}$ and
$\varphi_{2}$ are the conduction electron 
states corresponding to d-electron states possessing
$0$, $\uparrow$, $\downarrow$ and $2$ electrons, respectively.
In the Anderson Hamiltonian, the energy gain comes from the mixing term, 
while the Coulomb repulsion increases the energy. Therefore, 
the expectation value of the mixing term should have finite value 
in the ground state. This is the necessary condition for the ground 
state. If we assume $V_{dk}=Vu_k$, we obtain
\begin{eqnarray}
<\Psi_g|H_{mix}|\Psi_g>\neq 0, 
\end{eqnarray}
which is equivalent to 
\begin{eqnarray}
V\sum_{\sigma}[<A_0\varphi_0|\frac{1}{\sqrt{N}}\sum_k {u_k}^* 
C_{k\sigma}^{\dagger}|A_{\sigma}\varphi_{\sigma}> \nonumber \\
+<A_2\varphi_2|\frac{1}{\sqrt{N}}\sum_k u_k C_{k\sigma}
|A_{-\sigma}\varphi_{-\sigma}>\nonumber \\
+<A_{\sigma}\varphi_{\sigma}|\frac{1}{\sqrt{N}}\sum_k u_k 
C_{k\sigma}|A_0\varphi_0> \nonumber \\
+<A_{-\sigma}\varphi_{-\sigma}|\frac{1}{\sqrt{N}}\sum_k {u_k}^* 
C_{k\sigma}^{\dagger}|A_2\varphi_2>]\neq 0.
\end{eqnarray}
Here we define creation and annihilation operators of electron 
at the origin as
\begin{eqnarray}
C_{0\sigma}=\frac{1}{\sqrt{N}}\sum_k u_kC_{k\sigma}, \\
C_{0\sigma}^{\dagger}=\frac{1}{\sqrt{N}}\sum_k {u_k}^*C_{k\sigma}^{\dagger}.
\end{eqnarray}
In order for the above matrix elements not to vanish,
%it is necessary for the matrices to conserve 
the local electron numbers have to be conserved. From this condition,
 we can derive the following relations among conduction electron numbers 
associated to the i-component of d-electron states. Denoting the 
number of conduction electrons with $\sigma$ as $n_{i}^{\sigma}$, 
we obtain 
\begin{eqnarray}
n_{0}^{\sigma}=n_{\sigma}^{\sigma} +1,\\
n_{0}^{-\sigma}=n_{\sigma}^{-\sigma}=n_{-\sigma}^{\sigma}=n_{0}^{\sigma},\\
n_{2}^{\sigma}=n_{-\sigma}^{\sigma}-1,\\
n_{2}^{-\sigma}=n_{-\sigma}^{-\sigma}=n_{\sigma}^{\sigma}
\end{eqnarray}
Finally we obtain
\begin{equation}
n_{\sigma}^{\sigma}=n_{-\sigma}^{\sigma}-1
\end{equation}
Thus, we have the same relation as $(\ref{eqn:KonAn})$ for the 
s-d Hamiltonian. This fact means that the energy gain in the 
spin singlet state comes from the mixing term in the Anderson 
Hamiltonian. This mixing energy exists even in the case of $U=0$. 
The effect of $U$ for the s-d Hamiltonian is only suppressing the 
mixing term and reducing the energy gain.

Now we extend the above discussion to periodic systems such as 
the Hubbard Hamiltonian and the periodic Anderson Hamiltonian. 
In this case we assign the d-orbit in the impurity model to a 
lattice point chosen arbitrarily in the periodic system. 
We can write the ground-state wave function $\Psi_g$ as 
$(\ref{eqn:wavfun})$. Here the conduction electrons are 
replaced by local electrons around the site 0 which do not 
include 0-site. For the electron at a site and electrons 
around the lattice point, the relation shown in Fig.$\ref{fig:S1}$ 
should hold. In the the periodic Anderson Hamiltonian, 
the matrix element which should be finite is the mixing term. 
In the case of Hubbard Hamiltonian, transfer matrices to 
neighboring sites should not vanish. In these models the energy 
gain originates from the coherent mixing and transfer terms. 
That is, the coherent band-like motion is the origin of energy gain. 
For the mechanism to be possible, " conservation of local electron 
number" should be satisfied. This condition shows that the electron 
state at a lattice point always couples in the singlet state with 
electrons existing neighboring sites. 
This state is nothing but the resonating-valence-bond (RVB) state. 
The importance of this state is stressed by Anderson
 \cite{RVB}. 
The singlet state is the same as the singlet in the single impurity. 
Thus, we can understand that the RVB state in metals is nothing but 
the Fermi liquid state, although Anderson denies the Fermi liquid 
state as the normal state in HTSC and looks for RVB state.
%This is hopeless and contradictory effort. 
%We should stand on the Fermi liquid state. 
The fact " The Fermi liquid is a RVB state in metals" is the important 
result obtained from the study of the Kondo effect. The Fermi liquid 
state is the linear combination of local spin singlet states. 
Let us consider to shift the electron number uniformly in Fig.$\ref{fig:S1}$.
By the shift, the electronic states change 
from $(a)$ to $(b)$. After shifting the electron number, we can see 
that when a spin of electron is up at a lattice site, the neighboring 
electron number of this site possesses one up spin defect. That is, 
when an electron transfers between lattice sites, the local spin 
conservation holds always. The Hamiltonian conserves the total spins. 
In the ground state, spin conservation holds also locally. 
The local conservation of spin is a stronger condition than the 
total conservation. This result means when we need to bring spin 
from far distant region, a coherent electron motion is impossible. 
An electron possessing a spin can transfer by accompanying the same 
spin hole around it, and the system keeps the uniform state in spin 
distribution as a whole. 
When we separate the wave-function into their local components, the same 
singlet state as in the Kondo effect is realized. This point is 
very important. For the heavy fermion system, the same argument 
is applicable by using the periodic Anderson Hamiltonian.\cite{YYHan}

{\it 2-2.Perturbation expansion for the Anderson Hamiltonian}\\
We use the following perturbation expansion with respect to $U$: 
\cite{KYamada2}
\begin{eqnarray}
H&=&H_0+H',
\label{eqn:Per}
\end{eqnarray}
where
\begin{eqnarray}
H_0&=&\sum_{k\sigma}\varepsilon_k C_{k\sigma}^{\dagger}C_{k\sigma}
+\frac{1}{\sqrt{N}}\sum_{k\sigma}(V_{kd}C_{k\sigma}^{\dagger}
d_{\sigma}+V_{dk}d_{\sigma}^{\dagger}C_{k\sigma})\nonumber \\
& &+\sum_{\sigma}E_{d\sigma}n_{d\sigma}-U<n_{d\uparrow}><n_{d\downarrow}>,\\
E_{d\sigma}&=&\varepsilon_d+U<n_{d-\sigma}>,\\
H'&=&U(n_{d\uparrow}-<n_{d\uparrow}>)(n_{d\downarrow}-<n_{d\downarrow}>).
\end{eqnarray}
We expand the fluctuation term which is neglected in the Hartree-Fock 
approximation, assuming that $<n_{d\sigma}>=1/2$ and the expansion 
parameter is $u=U/(\pi\Delta)$. Then, we find that the ground state 
energy is given by 
\begin{eqnarray}
E_g=E(u=0)+\pi\Delta(-\frac{u}{4}-0.0369u^2+0.0008u^4+\cdots), 
\end{eqnarray}
%%%%%%%%%%%%%%%%%%%%%%%%%%%%%%%%%%%%%%%%%%%%%%%%%%%%%%%
\begin{figure}
\begin{center}
\epsfig{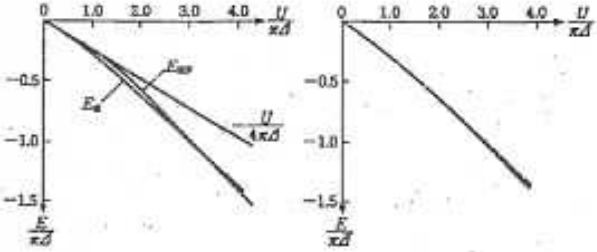}
\end{center}
\caption{
The ground state energy obtained by perturbation expansion up to 
the fourth order term.
The right figure shows the ground state energy based on the exact solution 
by Kawakami and Okiji (full line) and the result obtained by 
perturbation calculation(dotted line)}
 [Reference \cite{NKAO1}].
  \label{fig:Ag}
\end{figure}
%%%%%%%%%%%%%%%%%%%%%%%%%%%%%%%%%%%%%%%%%%%%%%%%%%%%%%
This result is shown in Fig.$\ref{fig:Ag}$. The right figure shows 
the result with that of the exact solution obtained by Kawakami 
and Okiji. The perturbation theory is reliable up to $u=4$.

\begin{figure}
\begin{center}
\epsfig{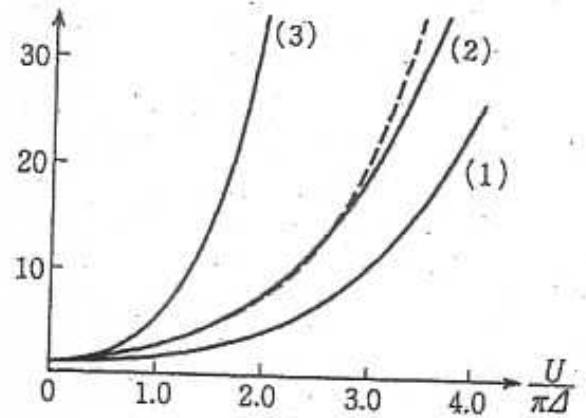}
\end{center}
\caption{
Results from perturbation calculation up to the fourth order terms.
$\tilde{\gamma},\tilde\chi_s, \tilde R
={\tilde{\chi}_{\uparrow\uparrow}}^2+{\tilde{\chi}_{\uparrow\downarrow}}^2/2$ 
are shown by (1),(2) and (3), respectively, 
in the function of $u=U/\pi\Delta$}
[References \cite{KYamada1}].
  \label{fig:Ap}
\end{figure}

Similarly we expand the spin susceptibility $\chi_s$, charge 
susceptibility $\chi_c$, and the electronic specific heat coefficient 
$\gamma$:
\begin{eqnarray}
\chi_s&=&\frac{(g\mu_B)^2}{2}\frac{1}{\pi\Delta}\tilde{\chi_s},\\
\label{eqn:chis}
\tilde{\chi_s}&=&\tilde{\chi}_{\uparrow\uparrow}
+\tilde{\chi}_{\uparrow\downarrow}=\sum_{n=0}a_nu^n.\\
\chi_c&=&\sum_{\sigma}\frac{1}{\pi\Delta}\tilde{\chi_c},\\
\label{eqn:chic}
\tilde{\chi_c}&=&\tilde{\chi}_{\uparrow\uparrow}
-\tilde{\chi}_{\uparrow\downarrow}=\sum_n a_n(-u)^n=\tilde{\chi_s}(-u),\\
\gamma&=&\frac{2\pi^2k_B^2}{3}\frac{1}{\pi\Delta}\tilde{\gamma},\\
\tilde{\gamma}&=&\frac{1}{2}(\tilde{\chi_s}+\tilde{\chi_c})
=\tilde\chi_{\uparrow\uparrow}.
\end{eqnarray}
Here, $\tilde{\chi}_{\uparrow\uparrow},\tilde{\chi}_{\uparrow\downarrow}$ 
are given by correlation functions between parallel and anti-parallel spins. 
The result obtained by perturbation calculation,
\begin{eqnarray}
\label{eqn:ky}
\tilde{\chi_s}&=&1+u+(3-\frac{{\pi}^2}{4})u^2+(15-\frac{3{\pi}^2}{2})u^3
+0.055u^4+\cdots .
\end{eqnarray}
The results are shown in Fig.$\ref{fig:Ap}$. i'P), (2) and (3) are  
$\tilde{\gamma},$  $\tilde{\chi_s}$ and  $\tilde{R}$, respectively. \\
%%%%%%%%%%%%%%%%%%%%%%%%%%%%%%%%%%%%%%%%%%%%%%%%%%%%%%%
%%%%%%%%%%%%%%%%%%%%%%%%%%%%%%%%%%%%%%%%%%%%%%%%%%%%%%
Finally we explain the result of the resistivity. The resistivity 
decreases in terms of $T^2$ with increasing temperature from 
unitarity limit which corresponds to the phase shift $\delta=\pm \pi/2$.
\cite{YosidaYamada}. The result is given by
\begin{eqnarray}
R=R_0[1-\frac{{\pi}^2}{3}(\frac{k_B T}{\Delta})^2
(2{\tilde{\chi}_{\uparrow\uparrow}}^2
+{\tilde{\chi}_{\uparrow\downarrow}}^2)+ \cdots ].
\end{eqnarray}
When $u$ is large, we can use the s-d Hamiltonian. By the relation, 
$\tilde{\chi}_{\uparrow\uparrow}=\tilde{\chi}_{\uparrow\downarrow}
=\tilde{\chi}_s/2$ we obtain 
\begin{eqnarray}
R=R_0[1-\frac{{\tilde{\chi}_s}^2}{4}(\frac{\pi k_B T}{\Delta})^2+ \cdots].
\end{eqnarray}
The $T^2$ dependence can be scaled by Kondo temperature $T_K$. 
In that case $T_K=2\Delta/\pi\tilde{\chi}_s$.
\\ \\
{\it 3. Continuity and Exact Solution}\\
In 1980, the exact solution for the s-d Hamiltonian on the basis 
of the Bethe hypothesis was obtained by Andrei\cite{Andrei} and 
Wiegmann\cite{Wiegmann}. For the Anderson Hamiltonian, Wiegmann 
\cite{PBWieg}, and Kawakami and Okiji \cite{NKAO1} obtained the 
exact solution. The results by the exact solution are the following;
\begin{eqnarray}
\tilde{\chi_s}&=&\sqrt{\frac{\pi}{2u}}\exp[\frac{{\pi}^2}{8}
-\frac{1}{2u}] 
+\frac{1}{\sqrt{2\pi u}}\int_{-\infty}^{\infty}
\frac{\exp[-x^2/2u]}{1+(\frac{\pi u}{2}+ix)^2}dx , \nonumber\\ 
\tilde{\chi_c}&=&\frac{1}{\sqrt{2\pi u}}
\int_{-\infty}^{\infty}\frac{\exp[-x^2/2u]}{1+(\frac{\pi u}{2}+x)^2}dx .
\end{eqnarray}@
The first term of $\chi_s$ seems to possess a singular term $\exp[-1/u]$ 
in $u$. However, Zlati$\acute{\rm c} $ and Horvati$\acute{\rm c}$ 
proved that the result is analytic. \cite{ZLH}
By using
\begin{eqnarray}
\tilde{\chi_s}&=&\exp[\pi^2 u/8]\sqrt{\frac{2}{\pi u}}
\int_0^{\infty}\exp[-x^2/2u]\frac{\cos(\pi x/2)}{1-x^2}dx ,\\
\tilde{\chi_c}&=&\exp[-\pi^2 u/8]\sqrt{\frac{2}{\pi u}}
\int_0^{\infty}\exp[-x^2/2u]\frac{\cosh(\pi x/2)}{1+x^2}dx , \nonumber\\ 
\end{eqnarray}
we can obtain the expansion coefficients of $(\ref{eqn:chis})$ 
and $(\ref{eqn:chic})$. The result is given by the following formula.
\begin{eqnarray}
a_n=(2n-1)a_{n-1}-(\frac{\pi}{2})^2a_{n-2} ,\\
a_0=a_1=1 ,
\end{eqnarray}
from which we obtain the solution,
\begin{eqnarray}
a_n=[(\frac{\pi}{2})^{2n+1}/(2n+1)!!]P_n ,\\
P_n=\sum_{k=0}^{\infty}\frac{(-1)^k}{k!}\frac{(2n+1)!!}{[2(n+k)+1]!!}
(\frac{{\pi}^2}{8})^k .
\end{eqnarray}
Here $P_n$ has the following properties,
\begin{eqnarray}
\frac{2}{\pi}=P_0< P_n< P_{\infty}=1 .
\end{eqnarray}
When $n$ is large, $a_n$ is given by
\begin{eqnarray} 
a_n\simeq (\frac{\pi}{2})^{2n+1}/(2n+1)!! .
\end{eqnarray}
As the result we can see that the expansion with respect to $u$ 
converges with infinite convergence radius, $u=\infty$.

 The Wilson ratio is the ratio of the susceptibility to the specific 
heat coefficient, 
\begin{eqnarray}
R_W=(\frac{\chi_s}{{2\mu_B}^2})/(\gamma/\frac{2\pi^2}{3}{k_B}^2)
\nonumber \\
=\frac{\tilde{\chi_s}}{\tilde{\gamma}}
=\frac{2\tilde{\chi_s}}{\tilde{\chi_s}+\tilde{\chi_c}} .
\end{eqnarray}
As $u$ increases, the charge fluctuations are suppressed and  
$\chi_c$ approaches zero. In this case the Wilson ratio $R_W$ 
approaches from $1$ to $2$ with increasing $u$.
The normal state of the interacting Fermi system is connected 
continuously from the free Fermi gas with increasing interaction.
\cite{PWA},\cite{LDL}. As we have stressed here, the magnetic 
impurity system is a typical Fermi liquid. \cite{YANET}\\

%%%%%%%%%%%%%%%%%%%%%%%%%%%%%%%%%%%%%%%%%%%%%%%%%%%%%%%%%%%%%%%%%%%%%%%%%%%%%%
%%%%%%%%%%%%%%%%%%%%%%%%%%%%%%%%%%
{\it 4. transport phenomena in high-T$_{\rm c}$ cuprates
and other strongly correlated systems.}\\
%%%%%%%%%%%%%%%%%%%%%%%%%%%%%%%%%%
The Fermi liquid theory
has been developed by many authors
to explain various
electronic properties in strongly correlated
systems like heavy fermion systems,
organic superconductors and the
high-$T_{\rm c}$ superconductors (HTSC's)
 \cite{KYamada1,YANET}.
In the case of HTSC, 
almost all the physical quantities 
deviates from the conventional 
Fermi liquid behaviors observed in usual metals,
which are called the non-Fermi liquid behaviors.
Recent progress on the Fermi liquid theory
has succeeded in explaining various non-Fermi liquid behaviors
in HTSC in terms of the
Fermi liquid state with strong antiferromagnetic 
(AF) fluctuations
 \cite{KYamada1,YANET,Moriya-rev}.
Spin fluctuation theories like the
SCR theory
 \cite{Moriya-rev} 
and the fluctuation-exchange (FLEX) approximation
 \cite{Bickers}
can reproduce $1/T_1T\propto T^{-1}$ and $\rho\propto T$
in HTSC's, especially the correct superconducting (SC)
order parameter, $d_{x^2\mbox{-}y^2}$, as well as the
optimum transition temperature 
$T_{\rm c} \approx 100$K.

However, the
transport phenomena under the magnetic field in HTSC's,
like the Hall coefficient 
$R_{\rm H}=(\s_{xy}/\s_{xx}^2)/B_z$ and the
magnetoresistance (MR) $\Delta\rho/\rho_0$,
have sometimes been recognized
as the hallmark of the breakdown of the Fermi liquid state,
because they seem too anomalous to understand 
in terms of the Fermi liquid theory.
For example,
$|R_{\rm H}|$ increases as the temperature
decreases below $T_0\sim 700$K, and $|R_{\rm H}|\gg 1/ne$
($n$ being the electron filling number)
at lower temperatures
 \cite{Satoh}.
The sign of $R_{\rm H}$ is positive
in hole-doped systems like
YBa$_2$Cu$_3$O$_{7-\delta}$ (YBCO) and
La$_{2-\delta}$Sr$_\delta$CuO$_4$ (LSCO),
whereas it is negative in electron-doped systems
like Nd$_{2-\delta}$Ce$_\delta$CuO$_4$ (NCCO),
although their Fermi surfaces (FS's) 
observed in ARPES measurements are hole-like
 \cite{Armitage}.
As for the MR,
$\Delta\rho/\rho_0 \propto T^{-4}$
holds for a wide range of temperatures
 \cite{Kimura,Ando}.
Thus, the Kohler rule
derived by a simple relaxation time approximation (RTA)
in a single-band model,
$R_{\rm H}\propto$const. and
$\Delta\rho/\rho_0 \propto \rho_0^{-2}\ (\propto T^{-2}$
in HTSC),
are totally broken in HTSC's, although it is satisfied
in usual metals.

The breakdown of the Kohler rule 
suggests that both $R_{\rm H}$ and
$\Delta\rho/\rho_0$ in HTSC's are strongly enhanced 
at lower temperatures by the strong correlation effect.
In the RTA, the total current ${\vec J}_\k$ 
is assumed as the mean-free-path of the quasiparticle
${\vec v}_\k \tau_\k$,
where $\tau_\k$ is the relaxation time and
${\vec v}_\k$ is the quasiparticle velocity.
The breakdown of the Kohler rule
in HTSC cannot be reproduced satisfactorily
within the RTA even if one assume the
$\k$- and the energy-dependences of $\tau_\k$ arbitrarily;
therefore many people have considered that
the ground state of the HTSC is the 
non-Fermi liquid in which the concept of the
quasiparticle is not valid any more.
For example, Anderson considered that
the Tomonaga-Luttinger liquid state
with two-kinds of relaxation time, $\tau$ and $\tau_{\rm H}$,
is realized
 \cite{Anderson2}.
Thus, transport phenomenon has been one of the
most strong objection against the Fermi liquid picture
in HTSC's.

However, we stress that
the RTA is insufficient for strongly correlated systems
because the macroscopic conservation laws are broken.
In this section,
we study the transport phenomena in HTSC
based on Kubo's linear response theory,
by taking the vertex correction (VC) for current 
into account, following the conserving approximation
by Baym and Kadanoff
 \cite{Baym}.
In correlated electron systems
the total current ${\vec J}_\k$ is given by the
quasiparticle velocity ${\vec v}_\k ={\vec \nabla}\e_\k$
plus other dragged quasiparticles $\Delta{\vec J}_\k$
induced by interactions.
In the Landau-Fermi liquid theory,
$\Delta{\vec J}_\k$ is called (a kind of) the back-flow,
which is expressed as the
the current VC in the microscopic Fermi liquid theory.
$\Delta{\vec J}_\k$ is totally dropped in the RTA.
However, in various situations,
one should take the current VC in the conserving way
to avoid unphysical results
 \cite{Baym}.
In the present article,
we show that various anomalous
$T$-dependences of the transport phenomena
observed in HTSC's are caused in fact 
by the current VC's.

The pseudo-gap phenomena in under-doped
systems below $T^\ast\sim200$K
is also one of the most important problems in HTSC.
Since various anomalous phenomena
related to the pseudo-gap cannot
be explained by the spin-fluctuation theory,
the origin of the pseudo-gap phenomena
has been studied intensively for years.
Recent theoretical works 
using the FLEX+T-matrix theory
have showed that
the strong $d_{x^2\mbox{-}y^2}$-SC fluctuations 
induced by the AF fluctuations is a
significant candidate for the pseudo-gap
 \cite{YANET,KYamada1,Nagoya-rev}.
In the present article,
we study the transport phenomena
below $T^\ast$ using the FLEX+T-matrix theory with 
the current VC's, and show that 
various anomalous transport coefficients
are well reproduced in a unified way.
The present work strongly supports that
the pseudo-gap phenomena in under-doped HTSC's
are caused by the strong $d_{x^2\mbox{-}y^2}$-SC 
fluctuations as discussed in Refs.
 \cite{KYamada1,YANET,Nagoya-rev} in detail.

\vspace{5mm}
%%%%%%%%%%%%%%%%%%%%%%%%%%%%%%%%%%%%%
{\it 4-1. transport phenomena above $T^\ast$}\\
%%%%%%%%%%%%%%%%%%%%%%%%%%%%%%%%%%%%%
In the present subsection,
we explain the origin of anomalous transport phenomena
above the pseudo-gap temperature $T^\ast$,
where the spin-fluctuation theory is meaningful.
Based on the Kubo formula, 
Eliashberg derived the general expression for 
the DC-conductivity $\sigma_{\mu\nu}$
which is exact within the most divergent terms
with respect to $\gamma_\k^{-1}$.
By generalizing his work,
exact expressions for the Hall conductivity 
$\sigma_{xy}$
 \cite{Kohno-Hall}
and the magneto conductivity
$\Delta\sigma_{xx} \equiv \sigma_{xx}({\bf B})-\sigma_{xx}(0)$
 \cite{Kontani-MR}
have been derived.
These expressions contain the 
total current ${\vec J}_\k$,
which is composed of the 
irreducible four-point vertices ($\Gamma^I$).
By imposing the Ward identity
$\Gamma^I =\delta\Sigma/\delta G$ in the expressions,
we can calculate
the Hall coefficient and the magnetoresistance (MR)
so as to satisfy the conservation laws.

%%%%%%%%%%%%%%%%%%%%%%%%%%%%%%%%%%%%%%%%%%%%%%%%%%%%%%%
\begin{figure}
\begin{center}
\vspace{5mm}
\epsfig{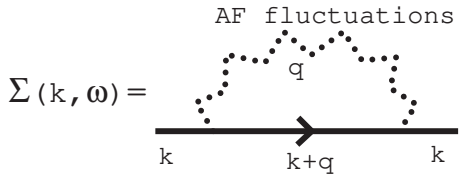}\\
\epsfig{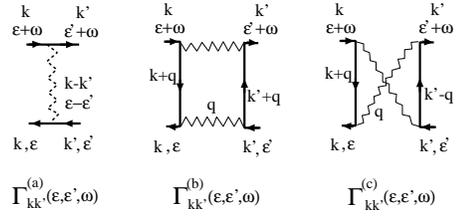}
\end{center}
\caption{
(a) Self-energy $\Sigma_\k(\w)$ 
 in the FLEX approximation.
(b) Irreducible four-point vertices 
 given by the Ward identity 
 $\Gamma^I=\delta\Sigma/\delta G$ 
 in the FLEX approximation.
}
  \label{fig:FLEX}
\end{figure}
\begin{figure}
\begin{center}
\vspace{5mm}
\epsfig{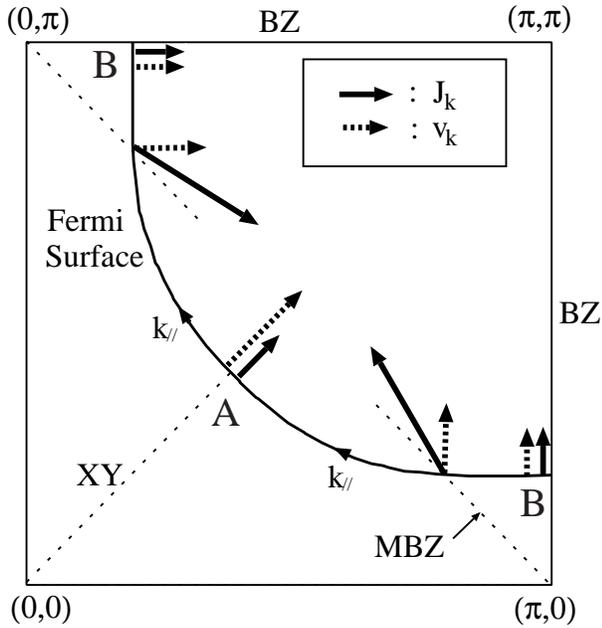}
\end{center}
\caption{
Schematic behavior of the total current
${\vec J}_\k$ and the quasiparticle velocity
${\vec v}_\k$ on the Fermi surface.
The cold spot is located around the 
point A (B) in the hole-doped (electron-doped) HTSC's}
[Reference \cite{Kontani-Hall}].
  \label{fig:J}
\end{figure}

%%%%%%%%%%%%%%%%%%%%%%%%%%%%%%%%%%%%%%%%%%%%%%%%%%%%%%
Here, we study 
various transport coefficients 
by using the FLEX approximation,
and taking the current VC's into account.
The self-energy in the FLEX approximation is 
given by the one-loop exchange
of the spin propagator in a self-consistent way,
as shown in Fig.\ref{fig:FLEX}.
The Bethe-Salpeter equation for the total current 
is given by
\begin{eqnarray}
{\vec J}_\k= {\vec v}_\k+\sum_{\k'}\int\frac{d\e}{4\pi i}
 {\cal T}_{\k\k'}^I(0,\e)|G_{\k'}^{\rm R}(\e)|^2{\vec J}_{\k'},
 \label{eqn:BS}
\end{eqnarray}
where ${\cal T}_{\k\k'}^I(0,\e)$ is 
given by the analytic continuation of
$\Gamma_{\k\k'}^I(\e_n,\e_{n'};\w_l)$ 
from the region $\e_{n}<0<\e_{n}+\w_l$
and $\e_{n'}<0<\e_{n'}+\w_l$
 \cite{Eliashberg}.
In the FLEX approximation,
${\cal T}_{\k\k'}^I$ is given by three diagrams,
which is shown in Fig.\ref{fig:FLEX}
 \cite{Kontani-Hall}.
Owing to the vertex correction,
${\vec J}_\k$ given by the solution of Eq.(\ref{eqn:BS})
is no more parallel to
the quasiparticle velocity ${\vec v}_\k$
when the AF fluctuations are prominent,
as explained in Ref.
\cite{Kontani-Hall} 
for the first time.
Apparently, the assumption in the RTA,
${\vec J}_\k=\tau_\k{\vec v}_\k$,
is invalid.
The schematic momentum dependence of ${\vec J}_\k$
is shown in Fig.\ref{fig:J}.
The physical reason for its anomalous $\k$-dependence
is that the total current behaves as 
${\vec J}_\k \propto {\vec v}_\k+{\vec v}_{\k+\Q}$
because the quasiparticle at $\k$
couples strongly to that at $\k+\Q$ through the current VC 
(i.e., the effective interaction between quasiparticles)
in the presence of the strong AF fluctuations.
The detailed explanation is given in Ref.
 \cite{Kontani-Hall}.
Note that the behavior of ${\vec J}_\k$
in Fig.\ref{fig:J} will be universal 
in nearly AF Fermi liquid beyond the FLEX
approximation once the Ward identity 
$\Gamma^I=\delta\Sigma/\delta G$ is satisfied
 \cite{Kontani-Hall}.

%%%%%%%%%%%%%%%%%%%%%%%%%%%%%%%%%%%%%%%%%%%%%%%%%%%%%%
%%%%%%%%%%%%%%%%%%%%%%%%%%%%%%%%%%%%%%%%%%%%%%%%%%%%%
Here, we discuss the Hall coefficient.
The general expression for $\sigma_{xy}$
is given by
 \cite{Kohno-Hall}
\begin{eqnarray}
\s_{xy}/B_z= -\frac{e^3}{4}\oint_{\rm FS}
 dk_\parallel |{\vec J}_\k|^2 \left(
 \frac{d\theta_\k^J}{dk_\parallel}\right) 
 \cdot \frac1{(\gamma_\k^2)} ,
 \label{eqn:s-xy}
\end{eqnarray}
where $\theta_\k^J=\tan^{-1}(J_{\k x}/J_{\k y})$.
$k_\parallel$ is the momentum parallel to the FS,
and $\gamma_\k=1/2\tau_\k={\rm Im}\Sigma_\k(-i\delta)$
represents the quasiparticle damping rate at $\k$.
As shown in Fig.\ref{fig:J},
$d\theta_\k^J/dk_\parallel$ is positive
around point A, whereas it is negative around point B.
According to the FLEX approximation,
the cold spot,
where $\gamma_\k$ takes the minimum value on the FS,
is located around point A(B) in hole-doped 
(electron-doped) systems
 \cite{Kontani-Hall}.
The cold spot locates at the most distant point
from the antiferromagnetic Brillouin zone boundary 
which connects ($\pi,0$) and ($0,\pi$).
Because the transport phenomena are governed mainly
by the quasiparticles around the cold spot,
it is theoretically expected that
$R_{\rm H}>0$ ($R_{\rm H}<0$)
in hole-doped (electron-doped) systems
when the AF fluctuations are strong.
We comment that the anisotropy of $\gamma_\k$
by the FLEX approximation is 
consistent with experiments by ARPES
 \cite{Armitage}.
The location of the cold spot in NCCO
was first predicted by the FLEX approximation
 \cite{Kontani-Hall},
and it was confirmed by ARPES later
 \cite{Armitage}.

\begin{figure}
\begin{center}
\vspace{5mm}
\epsfig{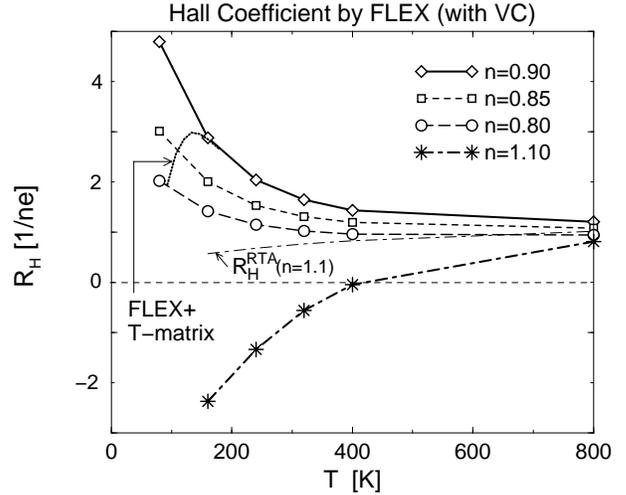}
\epsfig{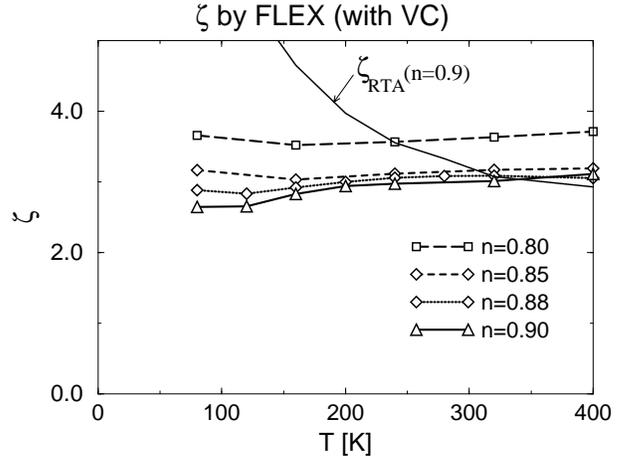}
\end{center}
\caption{
(i) $R_{\rm H}$ obtained by the FLEX approximation
which takes the current VC's into account.
$R_{\rm H}^{\rm RTA}$ by the RTA (without VC's)
shows only a weak $T$-dependence.
$R_{\rm H}$ by the FLEX+T-matrix approximation
is also shown; it starts to decrease 
below $T^\ast\sim150$K (pseudo-gap behavior).
(ii) $\zeta=\Delta\rho\cdot\rho/R_{\rm H}^2-1$
takes almost a constant value 
(the modified Kohler's rule)
by taking account of the current VC's.
On the other hand, $\zeta_{\rm RTA}$ shows
a strong $T$-dependence}
[H. Kontani: BUTSURI (Bulletin of the Physical Society of Japan)
{\bf 58} (2003) 524.].
  \label{fig:RH}
\end{figure}

Figure \ref{fig:RH} shows the
numerical result for $R_{\rm H}$ by the FLEX 
approximation, which takes the current VC into account.
The obtained $R_{\rm H}$ for hole-doped systems 
($n<1$) increases as the doping $\delta=|1-n|$ decreases.
On the other hand,
$R_{\rm H}$ for electron-doped systems ($n>1$)
changes to be negative below 500K.
These results are consistent with experiments.
We comment that
the negative $R_{\rm H}$ in electron-doped systems 
had been considered to
be an evidence of the non-Fermi liquid state 
because the curvature of the FS in HTSC
is positive everywhere, which directly means that
$R_{\rm H}>0$ within the RTA.
However,
it is naturally explained within the Fermi liquid picture
as the effect of the current VC
 \cite{Kontani-Hall}.
%%%%%%%%%%%%%%%%%%%%%%%%%%%%%%%%%%%%%%%%%%%%%%%%%%%%%
%%%%%%%%%%%%%%%%%%%%%%%%%%%%%%%%%%%%%%%%%%%%%%%%%%%%%%

The temperature dependence of $R_{\rm H}$
in the vicinity of the antiferromagnetic 
quantum-critical-point (AF-QCP) is 
\begin{eqnarray}
|R_{\rm H}|\propto \xi_{\rm AF}^2 ,
 \label{eqn:RH-scaling}
\end{eqnarray}
which can be derived from
the analytic consideration on the current VC
 \cite{Kontani-Hall}.
$\xi_{\rm AF}$ is the AF correlation length,
whose temperature dependence is
$\xi_{\rm AF}\propto T^{-1/2}$ 
according to the spin-fluctuation theory
(like the FLEX approximation or the SCR theory).
As a result,
the experimental fact $R_{\rm H}\propto T^{-1}$
in HTSC can be successfully reproduced in the present approach.

Similarly,
the MR due to the orbital motion of
conduction electrons
can be studied beyond the RTA,
based on the Kubo formula.
According to the analytic consideration
on the current VC,
we obtained the relation
 \cite{Kontani-MR}
\begin{eqnarray}
\Delta\rho/\rho_0
&\equiv& -\Delta\s_{xx}/\s_{xx}^0 - (\s_{xy}/\s_{xx}^0)^2
 \nonumber \\
&\propto& \xi_{\rm AF}^4\cdot \rho_0^{-2} ,
 \label{eqn:MR-scaling}
\end{eqnarray}
which suggests that
the Kohler's rule 
$\Delta\rho/\rho_0 \propto \rho_0^{-2}$
is violated because $\xi_{\rm AF}^4 \propto T^{-2}$ holds
when the AF fluctuation is strong.
Equation (\ref{eqn:MR-scaling})
explains the experimental relation
$\Delta\rho/\rho_0 \propto T^{-4}$
observed in YBCO and LSCO where $\rho_0 \propto T$.
Moreover,
Eqs.(\ref{eqn:RH-scaling}) and (\ref{eqn:MR-scaling})
directly mean the ``modified Kohler's rule''
 \cite{Kontani-MR-HTSC}
\begin{eqnarray}
\Delta\rho\cdot\rho_0 \propto R_{\rm H}^2 ,
\end{eqnarray}
which is well satisfied in various HTSC compounds
although the conventional Kohler's rule is completely broken
 \cite{Kimura,Ando};
therefore its reason has been discussed intensively
by many authors.
We stress that
it is naturally explained in the present study.
Figure \ref{fig:RH}
shows a numerical result by the FLEX approximation
including the current VC's.
The coefficient $\zeta=\Delta\rho\cdot\rho_0/R_{\rm H}^2-1$
is almost independent for a wide range of temperatures,
only when the VC's are taken into account correctly.
The experimental value 
is $\zeta=2\sim3$ in YBCO and Tl-based HTSC, and 
$\zeta\sim 13$ in optimally-doped LSCO,
which are well reproduced by the present study
which takes into account the difference of the curvature of the FS's
 \cite{Kontani-MR-HTSC}.

In summary,
the violation of the Kohler's rule in HTSC
had been frequently considered as an evidence 
of the violation of the quasiparticle picture.
However,
the present study has solved this
long-standing problem
in terms of the nearly AF Fermi liquid,
by taking the current VC.
Moreover, we stress that
the anomalous enhancement of the 
thermoelectric power (TEP), $S$,
is also explained from the standpoint of 
the Fermi liquid with strong AF fluctuations
 \cite{Kontani-S}.
Experimentally,
$S$ is positive in hole-doped systems at lower temperatures,
whereas it is negative in electron-doped ones.
This fact can be reproduced by the FLEX approximation,
as a natural consequence of 
the strong $\e$-dependence of $\tau_\k(\e)$
and the difference of the position of the cold spot
 \cite{Kontani-S}.

\begin{figure}
\begin{center}
\vspace{5mm}
\epsfig{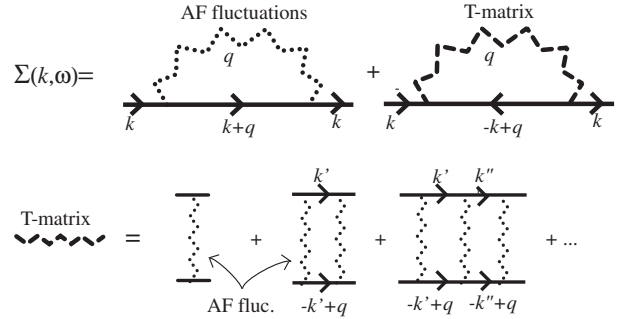}
\end{center}
\caption{
Self-energy $\Sigma_\k(\w)$ in the 
FLEX+T-matrix approximation.
The T-matrix for $d_{x^2\mbox{-}y^2}$-wave SC
is given by the ladder approximation 
for the AF fluctuations.
}
  \label{fig:FLEX-T}
\end{figure}

\vspace{5mm}
%%%%%%%%%%%%%%%%%%%%%%%%%%%%%%%%%%%%%
{\it 4-2. transport phenomena below $T^\ast$}\\
%%%%%%%%%%%%%%%%%%%%%%%%%%%%%%%%%%%%%
In under-doped HTSC's,
a deep pseudo-gap in the density of states 
around the chemical potential
emerges below $T^\ast \sim$200K,
which is called the pseudo-gap region.
The mechanism of the pseudo-gap
has been a central issue in HTSC for years.
Spin-fluctuation theories cannot reproduce it.
According to precise ARPES measurements,
the $\k$-dependence of the pseudo gap 
is equal to that of the SC gap function
with $d_{x^2\mbox{-}y^2}$-symmetry.
Moreover, it changes to the
SC gap below $T_{\rm c}$ smoothly
both in energy and in momentum.
This fact strongly suggests 
an  intimate relation between the pseudo-gap 
and the superconductivity.
Motivated by the experimental facts,
various strong coupling theories of the SC fluctuations
have been studied.
Several groups have developed 
the ``FLEX+T-matrix theory'';
it is equivalent to the FLEX approximation
at higher temperatures,
whereas the self-energy correction below $T^\ast$
by the strong SC fluctuations (T-matrix)
mediated by the AF fluctuations (FLEX)
is taken into account self-consistently
 \cite{KYamada1,YANET,Nagoya-rev}.
The self-energy in this approximation
is shown in Fig.\ref{fig:FLEX-T}.
The FLEX+T-matrix approximation
can reproduce the pseudo-gap in DOS,
$1/T_1T$, and the decrease of $T_{\rm c}$
in the under-doped region below $T^\ast$.
%%%%%%%%%%%%%%%%%%%%%%%%%%%%%%%%%%%%%%%%%%%%%%%%%%%%%%%
%%%%%%%%%%%%%%%%%%%%%%%%%%%%%%%%%%%%%%%%%%%%%%%%%%%%%%

\begin{figure}
\begin{center}
\vspace{5mm}
\epsfig{file=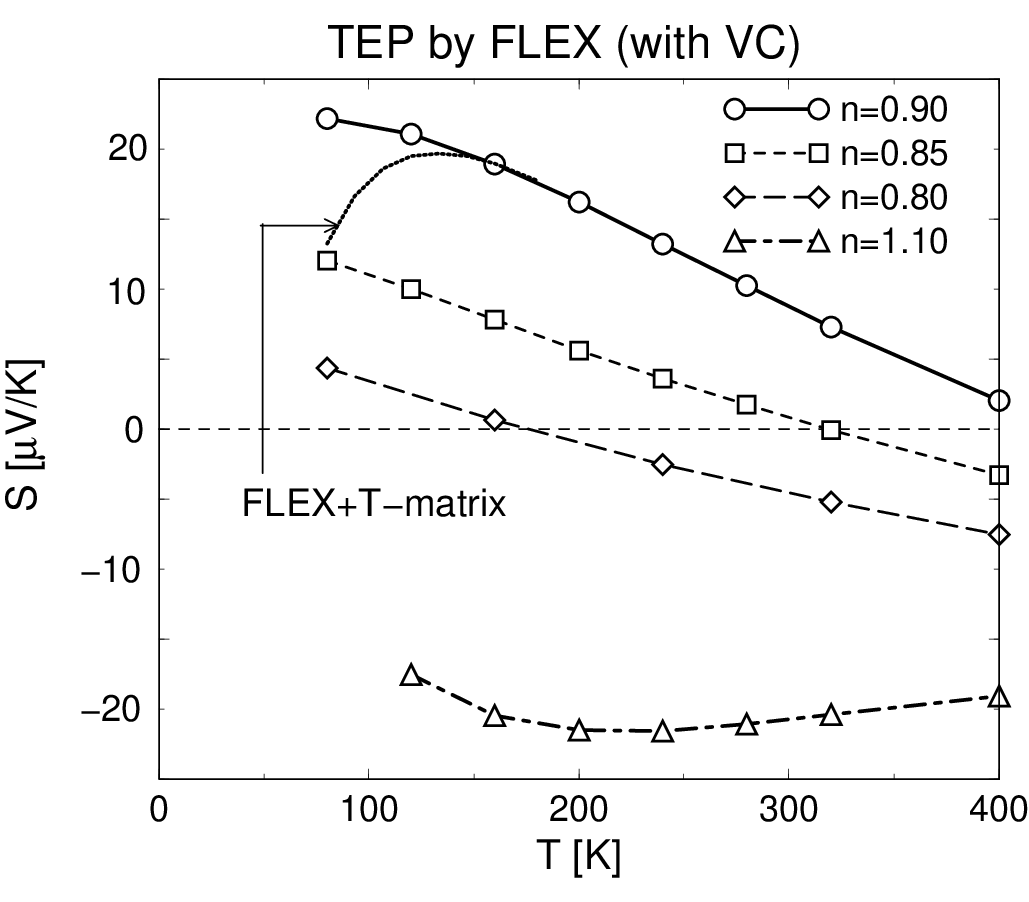,width=8cm}\\
\epsfig{file=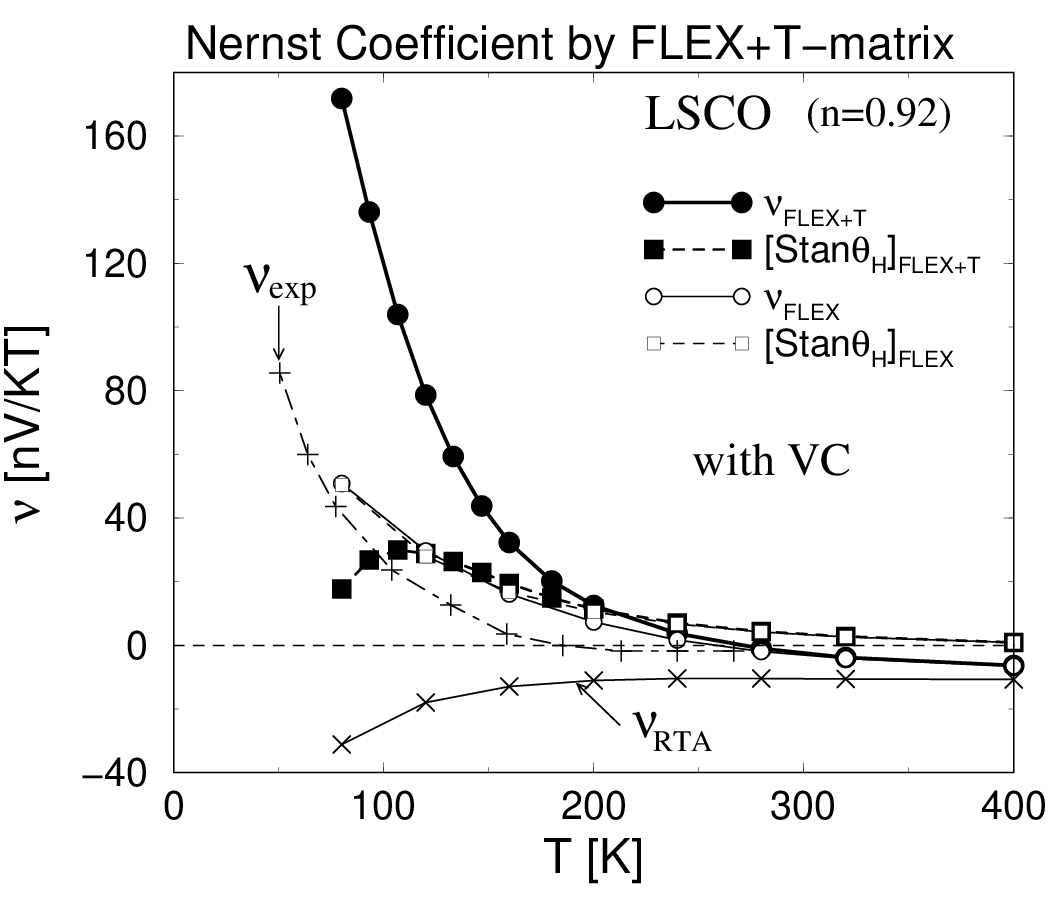,width=8cm}\\
\epsfig{file=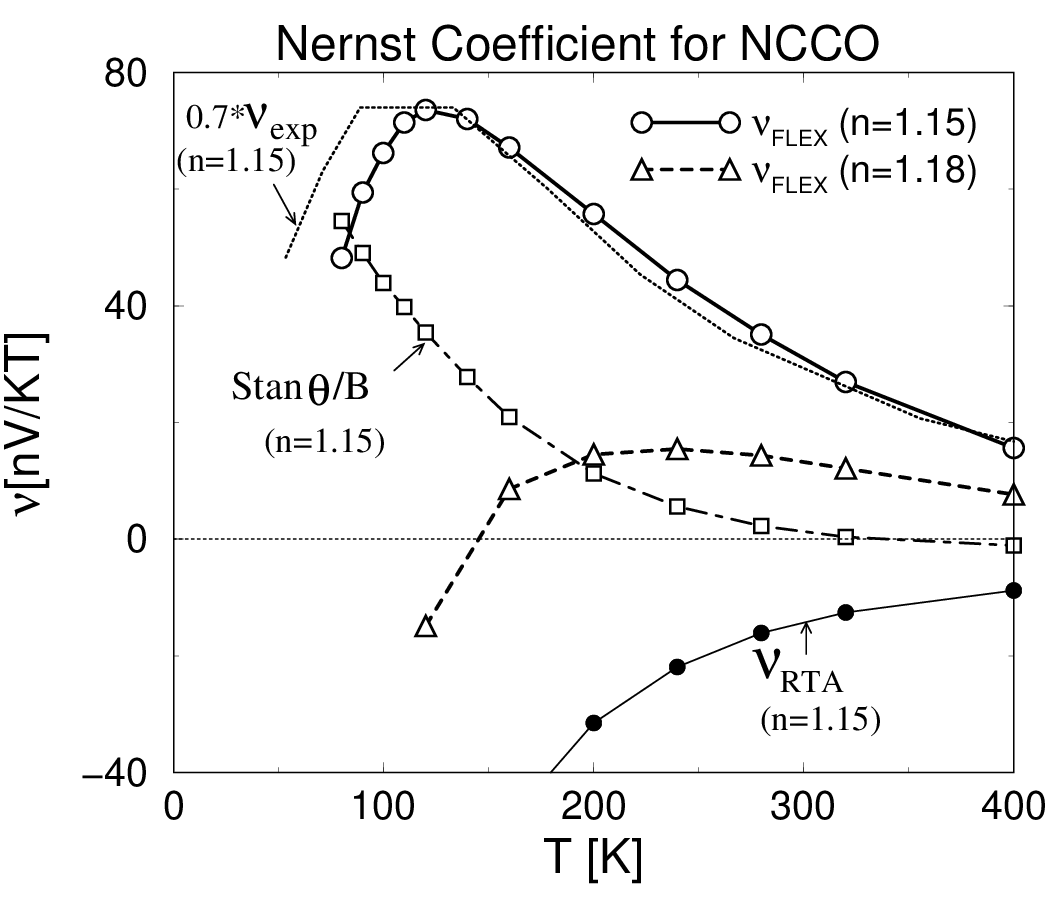,width=8cm}
\end{center}
\caption{}
(i) $S$ obtained by the FLEX+T-matrix approximation.
It starts to decrease below $T^\ast\sim150$K, which is
consistent with the experimental pseudo-gap behavior.
(ii)$\nu$ for LSCO obtained by the FLEX+T-matrix;
the abrupt increase of $\nu$ below $T^\ast$ is 
reproduced.
$\nu_{\rm exp}$ is cited from \cite{Ong}.
(iii)$\nu$ for NCCO obtained by the FLEX.
$\nu_{\rm exp}$ is cited from \cite{Fournier}
[H. Kontani: BUTSURI (Bulletin of the Physical Society of Japan)
{\bf 58} (2003) 524.].
  \label{fig:nu}
\end{figure}

It is possible to study 
various transport phenomena by the FLEX+T-matrix
approximation, taking the current VC's
as discussed in the previous section.
As shown in Figs.\ref{fig:RH} and \ref{fig:nu},
both $R_{\rm H}$ and $S$ start to decrease
below $T^\ast\sim$150K in the FLEX+T-matrix approximation;
it is consistent with experiments.
It is simply understood from the fact that
the AF fluctuations,
which gradually increase below $T_0=600\sim700$K,
start to decrease below $T^\ast$ reversely by reflecting the
pseudo-gap formation.
Because $R_{\rm H}$ and $S$ are strongly enhanced 
due to the AF fluctuations as discussed in the
previous section,
they decrease below $T^\ast$.

%%%%%%%%%%%%%%%%%%%%%%%%%%%%%%%%%%%%%%%%%%%%%%%%%%%%%%
%%%%%%%%%%%%%%%%%%%%%%%%%%%%%%%%%%%%%%%%%%%%%%%%%%%%%%
However, the Nernst coefficient $\nu$
($\nu\equiv S_{yx}/B_z = -E_y/B_z \nabla_x T$;
off-diagonal TEP under the magnetic field)
is strongly enhanced in the pseudo-gap region
about 100 times larger than the usual metals
 \cite{Ong,Capan}.
Such a nontrivial behavior of $\nu$
has attracted much attention as a key phenomenon 
closely related to the origin of the pseudo-gap.
According to the linear response theory,
$\nu$ is given by
\begin{eqnarray}
\nu= \left[ \a_{xy}/\s_{xx}^0 -S\s_{xy}/\s_{xx}^0 
 \right]/B_z,
 \label{eqn:nu1}
\end{eqnarray}
where $\a_{xy}$ is the off-diagonal
Peltier conductivity, which is given by
\begin{eqnarray}
\a_{xy}&=&B_z\cdot \frac{e^2}{T}
 \sum_\k \int\frac{d\e}{2\pi} \left(\frac{\d f}{\d\e}\right)
|{\rm Im}G_\k^{\rm R}(\e)||G_\k^{\rm R}(\e)|^2
 \label{eqn:axy} \nonumber \\
& &\times |{\vec v}_\k(\e)| \gamma_\k(\e)A_\k(\e)
 \label{eqn:nu2}, \\
A_\k(\e)&=&\left({\vec Q}_\k(\e)\times
 \frac{\d}{\d k_\parallel}\left({\vec J}_\k(\e)/\gamma_\k(\e)
 \right) \right)_z, \\
{\vec Q}_\k(\e)&=& \e\cdot{\vec v}_\k+\sum_{\k'}\int\frac{d\e'}{4\pi i}
 {\cal T}_{\k\k'}^I(\e,\e')|G_{\k'}^{\rm R}(\e')|^2{\vec Q}_{\k'}(\e'),
 \label{eqn:A}
\end{eqnarray}
where ${\vec Q}_\k(\e)$ is the total heat current
 \cite{Kontani-nu,Kontani-nu-HTSC}.
$\nu$ vanishes in a system with the free-electron
like dispersion $\e_\k= \k^2/2m$
because the first and the second terms
in Eq.(\ref{eqn:nu1}) cancel out with each other
within the relaxation time approximation,
which is known as the Sommerfeld cancellation.
In usual metals, $\nu$ is small
because of an approximate Sommerfeld cancellation.
However, 
$\nu$ takes an enhanced value in HTSC
because $\a_{xy}$ is much enhanced due to 
the current VC's given by both the AF- and SC-fluctuations; 
thus the Sommerfeld cancellation is totally broken below $T^\ast$,
as we will see below.

It is known that
$\nu$ takes a huge value
in the vortex-liquid state above $H_{\rm c1}$
in a clean two dimensional sample,
reflecting the high mobility of vortices.
In fact, $\nu$ is frequently used as a
sensitive probe for the mixed state.
Based on his observation,
Ong {\it et al.} claimed that
spontaneous vortex-antivortex pairs
emerge in under-doped systems below $T^\ast$,
and they govern the transport phenomena 
in the pseudo-gap region.
 \cite{Ong}.
However, his assumption contradicts with
other transport coefficients; for example,
flux-flow resistance does not appear below $T^\ast$.

\begin{figure}
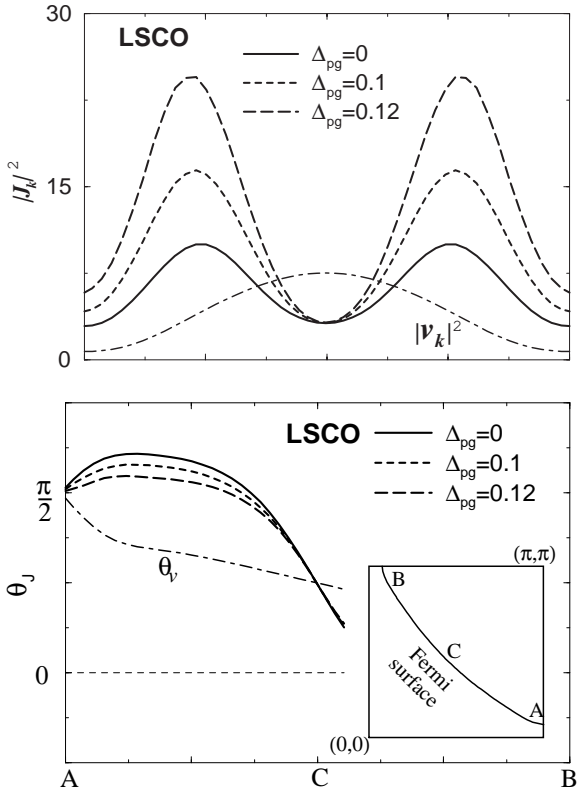

\begin{center}
\vspace{5mm}
\epsfig{file=Jabs-LSCO-new.eps,width=8cm}
\epsfig{file=Jang-LSCO.eps,width=8cm}
\end{center}
\caption{
$|J_\k|^2$ and $\theta_\k^J= \tan^{-1}(J_{\k x}/J_{\k y})$
given by the FLEX+T-matrix approximation, taking account
of the VC's due to AF- and SC-fluctuations.
The anisotropy of $|J_\k|$ is enhanced
as the pseudo-gap ($\Delta_{\rm pg}$) increases}
[Reference \cite{Kontani-nu-HTSC}].
  \label{fig:J2}
\end{figure}

Based on the FLEX+T-matrix approximation
with current VC's,
such an anomalous behavior of $\nu$ 
can be naturally understood as quasiparticle
transport phenomena, without assuming any
vortex-like excitations
 \cite{Kontani-nu-HTSC}.
The enhancement of $\nu$ below $T^\ast$
comes from the following facts
 \cite{Kontani-nu-HTSC}:
(i) The total current ${\vec J}_\k$ is strongly enhanced
by the Maki-Thompson type VC's by SC fluctuations,
except for the nodal direction of the SC gap, i.e.,
$(\pm\pi,\pm\pi)$-direction for the 
$d_{x^2\mbox{-}y^2}$-wave SC state.
(ii) Due to the current VC's by the AF-fluctuations,
${\vec J}_\k$ is not parallel to the 
total heat current ${\vec Q}_\k$ 
because the effect of the VC on ${\vec Q}_\k$ is 
not prominent in general, 
whereas ${\vec J}_\k \parallel {\vec Q}_\k$
is assumed (without any reason) in the RTA
 \cite{Kontani-nu,Kontani-nu-HTSC}.
The numerical result by the FLEX+T-matrix approximation
is shown in Fig.\ref{fig:J2}
 \cite{Kontani-nu}.
Although the facts (i) and (ii)
strongly enhance $\nu$ below $T^\ast$,
whereas (i) and (ii) influence $\rho$, $R_{\rm H}$ and $S$
only slightly, so they start to decrease moderately 
below $T^\ast$.

%%%%%%%%%%%%%%%%%%%%%%%%%%%%%%%%%%%%%%%%%%%%%%%%%%%%%%
%%%%%%%%%%%%%%%%%%%%%%%%%%%%%%%%%%%%%%%%%%%%%%%%%%%%%%
In conclusion,
we studied $\rho$, $R_{\rm H}$, $\Delta\rho/\rho_0$,
$S$ and $\nu$ both under and above $T^\ast$, 
using the FLEX+T-matrix approximation.
The results are shown in Figs.
\ref{fig:RH} and \ref{fig:nu}.
We found that 
various anomalous transport phenomena
in the pseudo-gap region ($T<T^\ast \sim 150$K in the present calculation)
are well reproduced in a unified way
in term of the Fermi liquid theory,
considering that the SC fluctuations
grow prominently in the pseudo-gap region.
Moreover, we stress that the huge Nernst signal
in electron-doped systems, 
which starts to increase below $\sim$300K 
in proportion to $T^{-1}$, is also 
reproduced by taking account of the current VC's
by taking only the AF-fluctuations.

Finally, we comment that
the MR is also enhanced further below $T^\ast$,
similarly to $\nu$.
It is also reproduced by the present work
if we take account of the VC for ${\vec J}_\k$
due to the SC fluctuations (see (i) listed above).
As shown in Ref.
\cite{Kontani-nu-HTSC},
$\Delta\rho\cdot\rho_0$ in the FLEX+T-matrix approximation
increases in proportion to $T^2$ above $T^\ast$,
and grows further below $T^\ast$.
It is consistent with available experiments.

\vspace{5mm}
%%%%%%%%%%%%%%%%%%%%%%%%%%%%%%%%%%%%%
{\it 4-3. $\kappa$-(BEDT-TTF)$_2$X, CeCoIn$_5$}\\
%%%%%%%%%%%%%%%%%%%%%%%%%%%%%%%%%%%%%
$\kappa$-(BEDT-TTF)$_2$X is a 
strongly correlated two-dimensional 
superconductor.
Its pressure-temperature (P-T) phase diagram,
which is often called the Kanoda phase diagram,
the SC phase in the higher pressure region
is close to the Mott (antiferromagnetic) 
insulating phase in the lower pressure region
 \cite{Kanoda}.
The maximum SC transition temperature
is $T_{\rm c}\approx$12K.
The simplest effective model
for $\kappa$-(BEDT-TTF)$_2$X is 
the anisotropic triangular Hubbard model
at half filling
 \cite{Kino-Fukuyama}.
By analysing this model 
with reasonable model parameters
based on the FLEX approximation,
the main feature of the P-T phase diagram 
is well reproduced
 \cite{Kino-Kontani,Kondo-Moriya,Schmalian}.
The symmetry of the SC state is 
$d_{x^2\mbox{-}y^2}$-wave
according to both the FLEX approximation
and the third-order perturbation theory
with respect to $U$
 \cite{KYamada1,YANET}.

The resistivity and the Hall coefficient
in $\kappa$-(BEDT-TTF)$_2$Cu[N(CN)$_2$]Cl
under $P=4.5\sim10$Kbar was measured in Ref.
\cite{Murata}.
In their measurement under high pressures,
the effect of the thermal contraction of the sample,
which is never negligible at ambient pressure,
is expected to be much reduced.
They found the approximate relations
$\rho \propto T$ and $R_{\rm H}\propto T^{-1}$
for $T=30\sim100$K.  Moreover, the Hall angle
$\cot \theta_{\rm H}\equiv \s_{xx}/\s_{xy}$
is proportional to $T^2$, as observed in
various HTSC's.
A similar behavior has recently been observed 
in $\kappa$-(BEDT-TTF)$_2$Cu[N(CN)$_2$]Br
by Taniguchi et al
 \cite{Taniguchi}.
These non-Fermi liquid behaviors of the
transport properties in $\kappa$-(BEDT-TTF)$_2$X
are well reproduced by the FLEX approximation
when the current VC's are taken into account
 \cite{Kontani-Kino}.

CeCoIn$_5$ is a quasi two-dimensional heavy 
fermion superconductor with $T_{\rm c}=2.3$K.
The symmetry of the SC is $d_{x^2\mbox{-}y^2}$-wave
according to the measurement of the
thermal conductivity under the magnetic field
 \cite{Izawa}.
It is consistent with the theoretical analysis
by the FLEX approximation and the 
perturbation theory with respect to $U$
 \cite{KYamada1,YANET}.
Transport phenomena in CeCoIn$_5$
are very remarkable
 \cite{Nakajima}:
Under 20K, $\rho\propto T$, $R_{\rm H} \propto T^{-1}$
and the modified Kohler's rule
$\Delta\rho\cdot\rho_0 \propto R_{\rm H}^2$ 
is satisfied quite well.
These behaviors, which are observed in 
various HTSC's, are expected to 
be reproduced theoretically if one take the 
current VC's caused by the strong AF fluctuations
in CeCoIn$_5$.
These anomalous transport phenomena,
which are expected to appear in nearly AF Fermi liquids,
confirm the fact that the spin fluctuations theory
can describe well electronic
properties in CeCoIn$_5$.

More interestingly,
the relations $\s_{xy}\propto B_z$ and 
$\Delta\s_{xx}\propto B_z^2$, both of which 
are assured by Onsager's reciprocal theorem
when $|B_z|$ is small enough,
are violated above the very low characteristic
fields $B_z^\ast$ ($\sim 0.1$Tesla),
although $\w_c^\ast\tau=(eB_z^\ast/mc)\tau\ll 1$
is expected to be satisfied.
These unexpected anomalous field dependences
of $\s_{xy}$ and $\Delta\s_{xx}$ will come
from the fact that CeCoIn$_5$ is 
very close to the AF-QCP:
According to current VC's in nearly AF metals,
relations 
$\s_{xy}\propto \xi_{\rm AF}^2 \tau^2 \cdot B_z$ and
$\Delta\s_{xx}\propto \xi_{\rm AF}^4 \tau^3 \cdot B_z^2$
are expected.
The field dependence of $\chi(Q,0)\propto \xi_{\rm AF}^2$,
which will be sensitive to the outer magnetic field
in the vicinity of the AF-QCP
 \cite{Sakurazawa},
should cause the deviation from
 $\s_{xy}\propto B_z$ and $\Delta\s_{xx}\propto B_z^2$.
In fact, experimentally, the ratio
$p\equiv (\rho_{xx}(B)-\rho_{xx}(0))/\s_{xy}^2(B)\rho_{xx}^2(0)
\ [\propto \xi_{\rm AF}^0 \tau^0]$
is nearly constant for $5\sim20$K and $0\sim4$Tesla,
although the usual Kohler's rule is completely broken
 \cite{Nakajima}.
This fact strongly supports that
anomalous transport phenomena in CeCoIn$_5$ 
can be well described by the current VC's caused by the
strong AF fluctuations.

The Nernst signal $\nu$ in CeCoIn$_5$ is also very anomalous
 \cite{Behnia}.
Below 20K,
it starts to decreases approximately
in proportion to $T^{-1}$,
taking an anomalously huge negative value
($\nu\sim-1\mu$V/KT) below 4K.
This behavior is very similar to the
Nernst signal in electron-doped HTSC;
it starts to increase below 300K
in proportion to $T^{-1}$, and it takes the
maximum value $\sim 0.1\mu$V/KT,
which is still about 100 times larger than
that in usual metals.
As we have shown above,
the temperature dependence and the 
magnitude of $\nu$ in NCCO are well reproduced by the current VC's
caused by the strong AF fluctuations,
even in the absence of the SC fluctuations.
(Note that the relation $\nu\propto T^{-1}$
is the hallmark of the importance of the AF fluctuations.)
Thus, it is naturally expected that
the giant Nernst signal in CeCoIn$_5$ is caused by 
the current VC's due to the AF fluctuations.
Considering that $\nu$ 
is proportional to $\tau \propto \rho_0^{-1}$,
experimental facts
$\nu_{\mbox{\tiny CeCoIn$_5$(4K)}}/\nu_{\tiny \mbox{NCCO(100K)}}
\approx 1(\mu$V/KT$)/0.1(\mu$V/KT$) =10$ and
$\rho_{\tiny \mbox{CeCoIn$_5$(4K)}}/\nu_{\tiny \mbox{NCCO(100K)}}
\approx 5(\mu\Omega)/50(\mu\Omega) =1/10$
strongly suggest that the origin of the
huge $\nu$ in CeCoIn$_5$ is the current VC's.

Finally, we comment on the effect of the SC fluctuations
on the Nernst effect.
In hole-doped HTSC's,
$\nu$ increases much faster than $T^{-1}$
below $T^\ast$, where the SC fluctuations are enhanced.
According to the study on 
the Nernst signal due to the quasiparticle contribution
by the FLEX+T-matrix theory,
the abrupt increase of $\nu$ in the pseudo-gap region 
is the cooperative phenomenon between the 
d-wave SC fluctuations and the AF fluctuations.
The SC fluctuations alone will not cause 
a prominent enhancement of $\nu$
because the total current ${\vec J}_\k$
due to the SC-fluctuations is parallel to ${\vec v}_\k$
(see Eqs. (\ref{eqn:axy}), (\ref{eqn:A}), and Ref. 
\cite{Kontani-nu-HTSC}).
%as far as the quasiparticle contribution is considered.

The effect of SC fluctuations on the
quasiparticle transport coefficients 
in the vicinity of $T_{\rm c}$,
which is called the Maki-Thompson term,
has been studied previously for weak-coupling superconductors.
The theory of transport phenomena developed by the authors 
based on the FLEX+T-matrix approximation
 \cite{Kontani-nu-HTSC}
is an extension of the theory of Maki-Thompson
in that (i) the cooperatation between the AF- and 
SC-fluctuations are considered, and (ii) the 
strong coupling situation where the inelastic 
quasiparticle damping due to AF- and SC-fluctuations
dominates the elastic one.
In addition, one have to solve the Bethe-Salpeter equation
(\ref{eqn:BS}) self-consistently in strongly correlated systems
to obtain reasonable results
 \cite{Kontani-Hall}.
In the case of HTSC's, 
the current VC's are playing important roles in various anomalous
transport phenomena which cannot be reproduced
within conventional Maki-Thompson terms.

On the other hand, 
the transport phenomena caused by 
the short-living Cooper pairs
(not by quasiparticles)
are called the Azlamazov-Larkin term.
More precisely, it is the transport phenomenon
due to the amplitude fluctuations of the
SC gap function, not by the vortex-like excitations. 
Recently,
the Nernst signal due to the Azlamazov-Larkin term
was studied in the particle-hole symmetric case
 \cite{Uss}.
The authors claimed that
the abrupt increment of $\nu$ in HTSC below $T^\ast$
can be explained by the Azlamazov-Larkin term.
However, 
the derived Azlamazov-Larkin term is 
$\nu\propto O(\tau^{-1})$,
whereas the quasiparticle transport term is
$\nu\propto O(\tau)$, that is,
the latter is $\tau^{2}$-times larger
than the former.
Thus, the Azlamazov-Larkin term for $\nu$
can exceed the quasiparticle term
only in the close vicinity of $T_{\rm c}$,
when the Sommerfeld cancellation is approximately realized.
As we have shown above, however, 
the Sommerfeld cancellation is totally broken
in HTSC's due to the current VC's.
In the under-doped HTSC's,
the diamagnetism due to the 
fluctuation of the SC order parameter above $T_{\rm c}$,
which is represented theoretically by the 
Azlamazov-Larkin term and is known to sensitive to
the strength of the field, is observed 
only in the vicinity of $T_{\rm c}$;
$T\simle T_{\rm c}+$5K
 \cite{Ong,Carballeira}.
In conclusion,
the abrupt increment of $\nu$
below $T^\ast$($\simge$100K) should be 
the quasiparticle origin, as discussed in Ref.
 \cite{Kontani-nu-HTSC}.

\vspace{5mm}
%%%%%%%%%%%%%%%%%%%%%%%%%%%%%%%%%%%%%
{\it 4-4. summary of this section}\\
%%%%%%%%%%%%%%%%%%%%%%%%%%%%%%%%%%%%%
We discussed the recent progress in
theoretical studies on the 
non-Fermi liquid like transport phenomena
in HTSC's, i.e.,
Hall coefficient, magnetoresistance,
electric thermopower and the Nernst coefficient.
They are well reproduced by the FLEX
or FLEX+T-matrix approximation 
in a unified way, if one takes the current VC's
into account.
In particular, anomalous transport phenomena
in the pseudo-gap region are understood 
successfully by the FLEX+T-matrix approximation,
which strongly supports that
the origin of the pseudo-gap in HTSC
is the strong SC fluctuations induced by
the AF fluctuations 
 \cite{KYamada1,YANET,Nagoya-rev}.
Quantitative numerical studies on the transport phenomena,
which takes into account the current VC's,
are at the beginning of progress.
We expect that fruitful results will be obtained in future
on the transport phenomena
in various strongly correlated electron systems,
not restricted to HTSC's.

\vspace{5mm}
%%%%%%%%%%%%%%%%%%%%%%%%%%%%%%%%%%%%%
{\it 5. Conclusions}\\
%%%%%%%%%%%%%%%%%%%%%%%%%%%%%%%%%%%%%
In this paper we have stressed that the spin singlet state discovered by 
Kondo effect is common to all the Fermi liquid irrespectively of impurity 
or periodic systems. Therefore, the resonating valence bond state in metals 
is nothing but the Fermi liquid state.
  When we study physical properties in strongly correlated electron 
systems, it is very important to consider the vertex correction terms to 
satisfy the conservation laws. In particular, since electron systems 
possess the lattice potential, the Umklapp scattering and spin fluctuation 
originating from electron interactions play the important roles in 
explaining  the anomalous behavior in terms of the Fermi liquid theory. We 
have shown the typical examples such as Hall coefficient, magnetoresistance 
and Nernst coefficient.
%%%%%%%%%%%%%%%%%%%%%%%%%%%%%%%%%%%%%%%%%%%%%%%%%%%%%%%%%%%%%%%%%%%%%

%%%%%%%%%%%%%%%%%%%%%%%%%%%%%%%%%%%%%%%%%%%%%%%%%%%%%%%%%%%%%%%%%%%%%

\end{multicols}{2}

\end{document}